\documentclass{webofc}
\usepackage[varg]{txfonts}   
\begin{document}
\title{Comparing two different descriptions of the I-V characteristic of graphene: theory and experiment}

\author{\firstname{Anatolii} \lastname{Panferov}\inst{1}\fnsep\thanks{\email{panferovad@info.sgu.ru}} \and
        \firstname{Stanislav} \lastname{Smolyansky}\inst{1,2} \and
        \firstname{David} \lastname{Blaschke}\inst{3,4,5} \and
        \firstname{Narine} \lastname{Gevorgyan}\inst{6} 
        }

\institute{Saratov State University, RU - 410026 Saratov, Russia
\and
	Tomsk State University, RU - 634050 Tomsk, Russia 
\and
	Institute for Theoretical Physics, University of Wroclaw, 50-204 Wroclaw, Poland
\and
	Bogoliubov Laboratory for Theoretical Physics, Joint Institute for Nuclear Research, RU - 141980 Dubna, Russia
\and
	National Research Nuclear University (MEPhI), RU - 115409 Moscow, Russia
\and
	Russian-Armenian University, 0051 Yerevan, Armenia
          }

\abstract{
The formalism of the nonperturbative description of transport phenomena in graphene on the framework of the quantum kinetic equation for the Schwinger-like process is compared with the description on the basis of Zener-Klein tunneling.
The regime of ballistic conductivity in a constant electric field is considered.

In the latter case the interaction of carriers with electric field is described in terms of the spatial dependence of their potential energy (x-representation).
The  presented kinetic formalism uses an alternative method of describing the interaction with a field through the introduction of a quasimomentum $P=p-(e/c)A(t)$ where $A(t)$ is the vector potential (t-representation).
Both approaches should lead to the same physical characteristics of the described process.

The measurement of the current in experiments is realized in static conditions determined by the potential difference between the electrodes and the distance between them.
These parameters are native for the x-representation.
On the contrary, in the approach based on the t-representation it is necessary to consider the situation in dynamics and  introduce the effective lifetime of the generated carriers.
In the ballistic regime this time depends on the distance between the electrodes.

We give a detailed comparison of these two descriptions of the current and demonstrate good coincidence with the experimental data of the alternative approach based on the t-representation.
It provides a reliable foundation for the application of nonperturbative methods adopted from strong field QED, that allows to include in the consideration more general models of the field (arbitrary polarization and time dependence) and to extend the scope of the theory.
}
\maketitle
\section{Introduction}
\label{intro}
In recent years considerable interest in a nonperturbative, dynamical description of transport phenomena in condensed matter physics inspired by the physics of strong electromagnetic fields (for example, \cite{Oka:2011ct} and references there).
Particular attention was devoted to graphene (see e.q., \cite{Yokomizo:2014sfa, Kao:2010hc}).
In this case there is an obvious similarity with the creation of electron-positron pair from vacuum in strong electromagnetic fields (Schwinger effect \cite{Schwinger:1951nm, fradkin, grib}).
In this context the nonperturbative kinetic approach \cite{Schmidt:1998vi, Blaschke:2008wf, Otto:2016pp, Blaschke:2017ax} belongs to the same class.
The analogous kinetic theory for graphene was constructed in the works \cite{bib_1, bib_2}.

The purpose of this work is a comparison of the results of the strong nonperturbative kinetic approach to graphene \cite{bib_1, bib_2} with the results of other works based on the WKB-type estimations and, in the end, with experiment.
As a test, the standard problem of calculation of the I-V characteristic for a rectangular graphene was chosen.
We want to show that in the case of a ballistic nondissipative regime, the nonperturbative kinetic approach leads to the same results in the case of a constant electric field as the methods developed in the works \cite{Vandecasteele:2010prb, Kane:2015, Dora:2010}.
This last approach is called commonly the Landauer approximation (or the Landauer-Datta-Lundstrom (LDL)\cite{Landauer:1957ibm, Landauer:1970pm, Landauer:1996mp}.
These approaches use the idea of the Zener-Klein (ZK) tunneling \cite{Zener:1934, Klein:1929} and can be used for calculations of the I-V characteristic (e.q., \cite{Datta:2012, Lundstrom:2013}).
Interaction of carriers with electric field in LDL approach is described in terms of the spatial dependence of their potential energy (x - representation).

The formalism presented in this paper \cite{bib_1, bib_2} uses an alternative method of describing the interaction with a field through the introduction of a quasimomentum $P=p-(e/c)A(t)$.
The characteristic of the field in this case is the vector potential $A(t)$.
Therefore, we will refer to this approach as the t - representation.

In the case of a constant electric field, the both approaches have to bring to identical observables.
But as the experimental conditions \cite{Vandecasteele:2010prb} for investigating the I-V characteristic in a constant electric field appropriate for a formulation of the theory in the x-representation, the problem arises of discussing the correspondence of results based on the x- and t-pictures in description of transport phenomena in graphene.
The analogical problem arises at an exact solution the task with a constant electric field in the t-representation \cite{Gavrilov:2017}.

\section{Kinetic equation approach \label{sect:ke}}

The basic kinetic equation (KE) for the description of electron-hole excitations in external electric fields with an arbitrary time dependence for the $D=2+1$ QED model of graphene has the form \cite{bib_1, bib_2}:
\begin{equation}
\dot{f}(\vec{p},t)=\frac{1}{2}\lambda \left( \vec{p},t\right)
\int\limits_{t_{0}}^{t}dt^{\prime }\lambda (\vec{p},t^{\prime })\left[ 1-2f(\vec{p},t^{\prime })\right] \cos \theta (t,t^{\prime }),  \label{KE}
\end{equation}
where parameter
\begin{equation}  
\label{lambda}
\lambda\left(\vec{p},t\right)=\frac{e v_{F}^{2}[E_{1}P_{2}-E_{2}P_{1}]}{\varepsilon^{2}(\vec{p},t)}
\end{equation}
is determined through the values of the components of the quasimomentum $P_{k}=p_{k}-\frac{e}{c}A_{k}(t)$, the energy of the quasiparticles $\varepsilon(\vec{p},t)=v_{F}\sqrt{P^{2}}= v_{F}\sqrt{(P_{1})^{2}+(P_{2})^{2}}$ and $v_{F} = 10^8~cm/s$ is the graphene Fermi velocity.
The phase in KE (\ref{KE}) is determined as
\begin{equation}
\theta (t,t^{\prime })=\frac{2}{\hbar }\int\limits_{t^{\prime
}}^{t}dt^{\prime \prime }\varepsilon (\vec{p},t^{\prime \prime }).
\label{phase}
\end{equation}

For the numerical analysis of the KE (\ref{KE}) for different field models it is appropriate to rewrite it in the form of an equivalent system of ordinary differential equations \cite{grib,Blaschke:2008wf}
\begin{equation}
\dot{f}=\frac{1}{2}\lambda u,\quad \dot{u}=\lambda \left( 1-2f\right) -\frac{2\varepsilon}{\hbar }
v ,\quad \dot{v}=\frac{2\varepsilon }{\hbar }u.
\label{KESyst}
\end{equation}
The carrier number density and the conduction current density are equal
\begin{eqnarray}
\label{density}
~~~~~~~~~~~~~~~~~~~~~~~~~n(t) & = & 8\int [dp]f(\vec{p},t), \\
\label{currentcomp}
~~~~~~~~~~~~~~~~~~~~~~~~~j_{i}^{\rm cond}(t) & = & 8e v^2_{F}\int [dp]  f(\vec p,t) \frac {P_i}{\varepsilon (\vec{p},t)} ,
\end{eqnarray}
where $[dp]=d^{2}p(2\pi \hbar )^{-2}$.
The coefficient $8$ takes into account two-valley degeneracy, two variants of the quasispin value and the equivalence of the contribution of quasielectrons and holes.
The auxiliary functions $u(\vec{p},t)$ and $v (\vec{p},t)$ describe polarization effects.

Let us consider the behavior of the solutions of the KE in a constant field, defining in the t-representation,
\begin{equation}
\label{fi1}
~~~E_1(t)= E_0 = {\rm const},~~A_1(t) = -E_0 t~; ~~~~~E_2(t) = 0, ~~A_2(t) = 0.  
\end{equation}

\begin{figure}[h]
\includegraphics[width=0.48\textwidth]{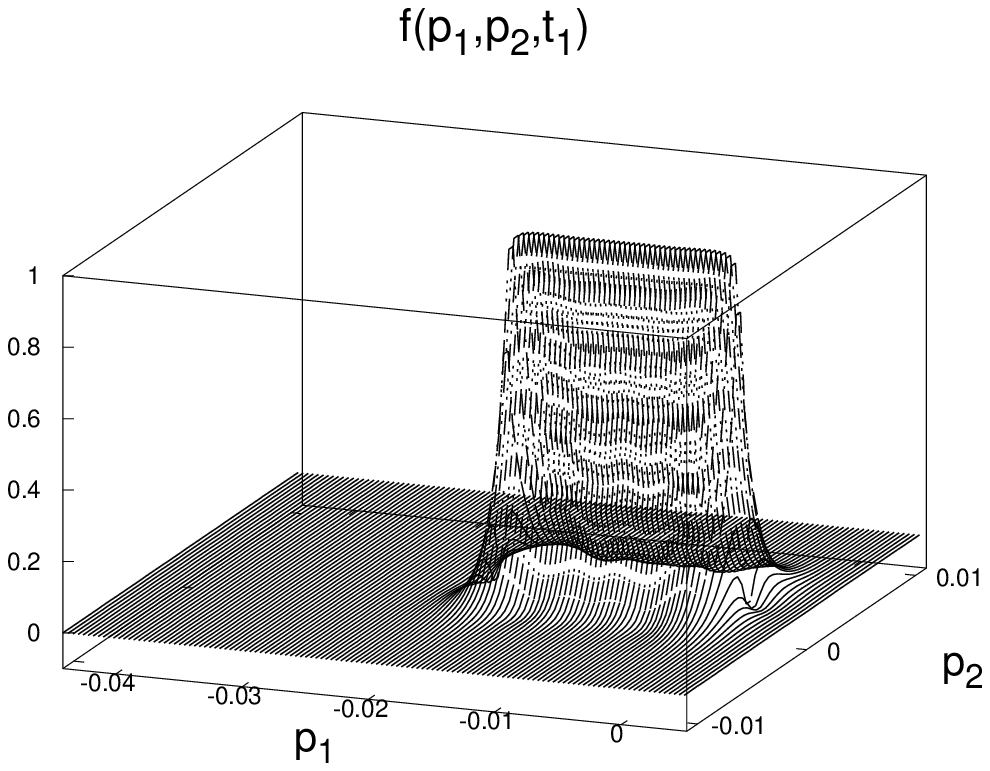} \hfill
\includegraphics[width=0.48\textwidth]{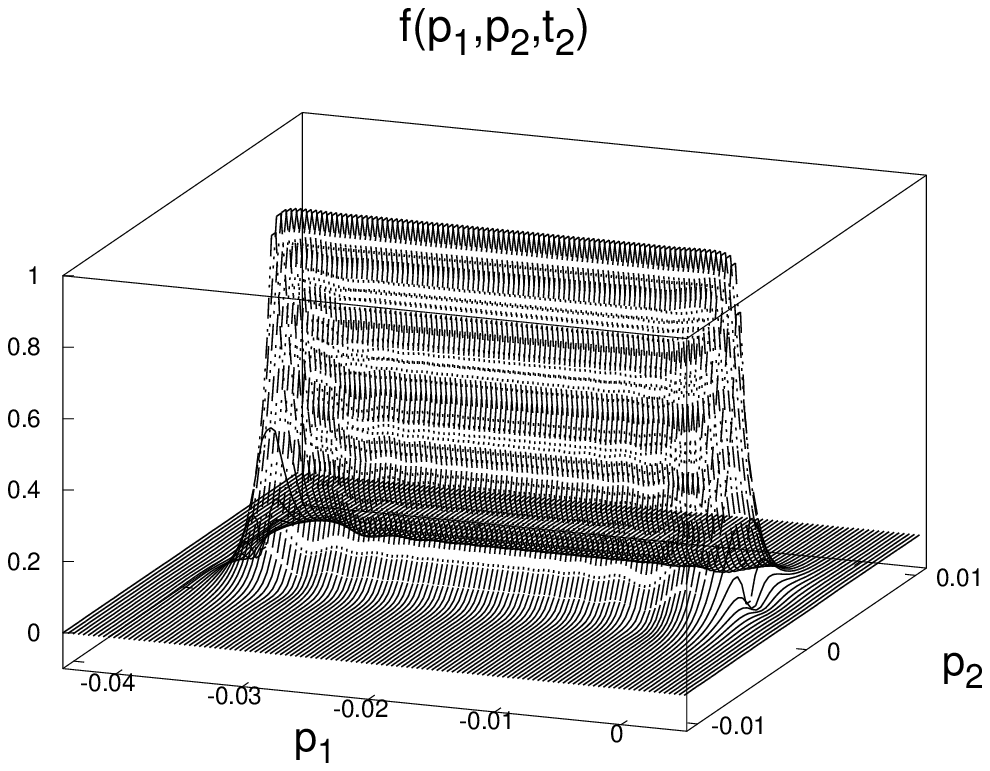}
\caption{The distribution function in the 2D momentum space for two consecutive points of time at the field strength $0.1~V/\mu m$.
{Left panel}: $t_1 = 0.5 \times 10^{-12}~s$,
{Right panel}: $t_2= 1.0 \times 10^{-12}~s$.
\label{fig:1}}
\end{figure}

\begin{figure}[h]
\includegraphics[width=0.48\textwidth]{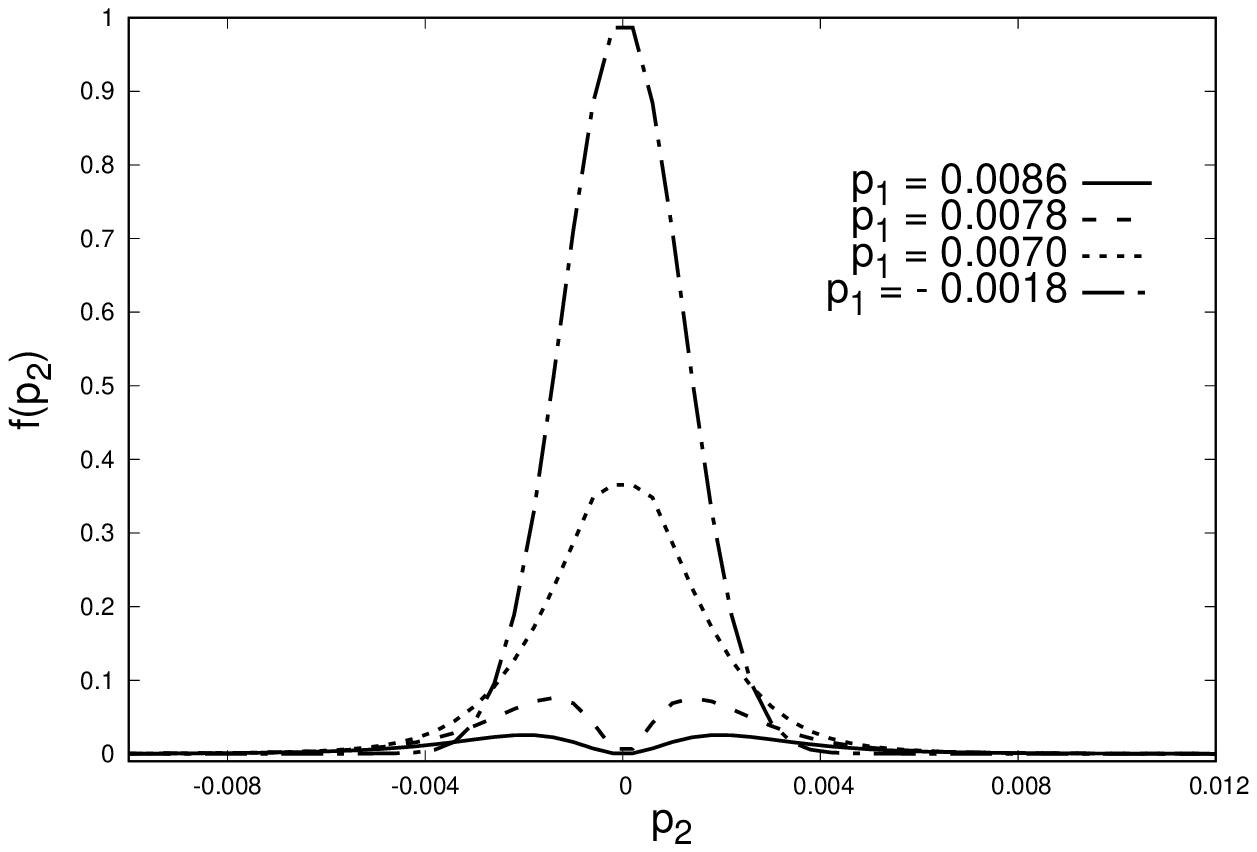} \hfill
\includegraphics[width=0.48\textwidth]{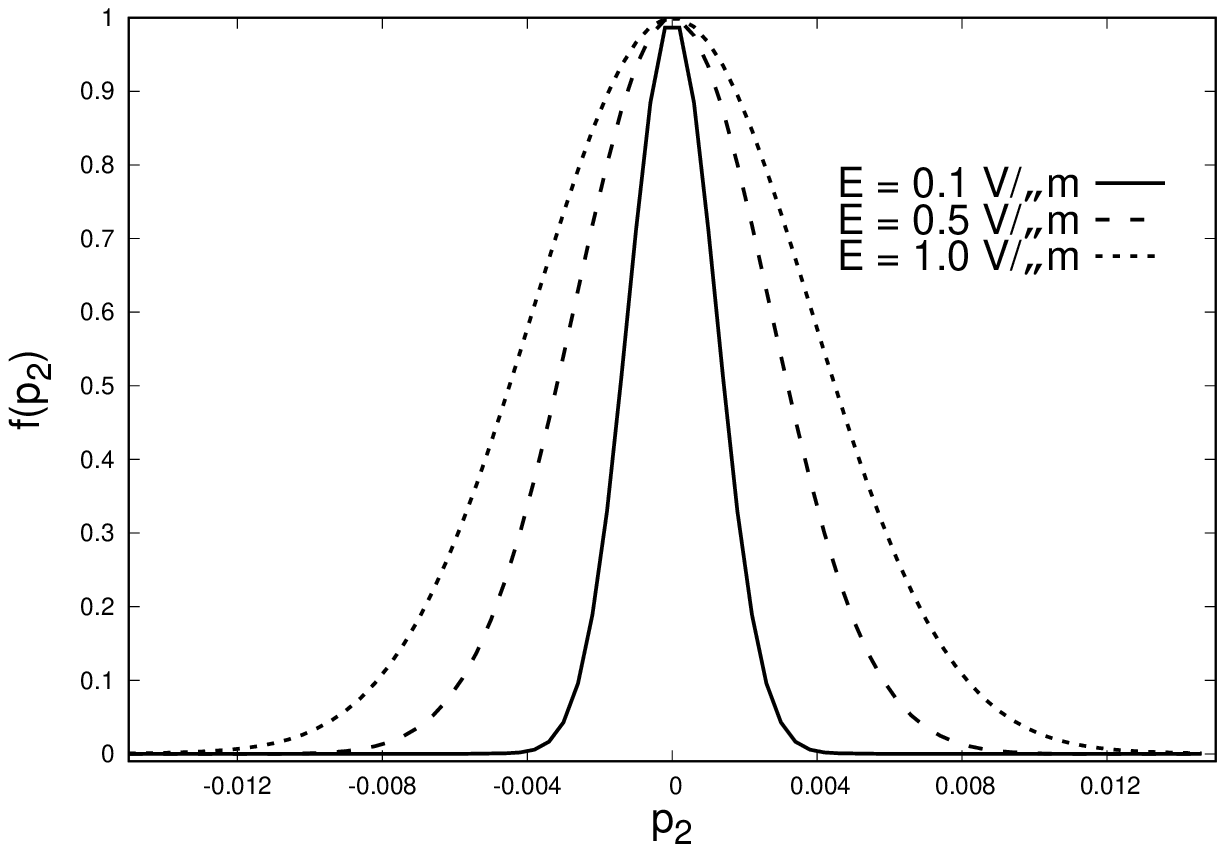}
\caption{
{Left panel}: The cross section $f(p_1,p_2)$ of the distribution function for small $|p_1|$ values,
{Right panel}:  The ``constant'' shape of the cross section $f(p_1,p_2)$ for $E = 0.1,~ 0.5~$ and $1.0~V/\mu m$.
\label{fig:5}}
\end{figure}

The Fig.~\ref{fig:1} shows the form of the distribution function for two consecutive points of time at a field strength $0.1~V/\mu m$.
The ``natural'' value $\hbar/a$ is used as a unit value for the $p_1, p_2$ ($a$ is the lattice constant).
Such kind accumulative behavior of the distribution function is in agreement with the results \cite{Fillion:2015} and \cite{Li:2018prb}.
The last work based on the Green function method takes into account the dissipative mechanism of inelastic scattering of optical phonons.

The distribution function of the excitation on the initial stage is formed and reaches the maximum value $f_{max}=1$ very rapidly.
Fig.~\ref{fig:5}, left panel, shows the transversal sections of this distribution for different values $p_1$ at early time in the  evolution, corresponding to the right edges of the distribution functions on Fig.~\ref{fig:1}.
In the case of the ballistic regime the following accumulation of population is a result of the increase of the longitudinal momentum $p_1$ in the direction of the acting electric field at invariable Gauss-like distribution under the transversal momentum $p_2$.
The halfwidth of the $p_2$-distribution is defined by the field strength $E$ (Fig.~\ref{fig:5}).
As a result, the carrier number density grows proportionally to the action time $T$ of the field, $n(T) \sim T$, as in the case of standard QED (e.q., \cite{grib}).

\section{LDL approximation \label{sect:ldl}}

In the LDL approach, the problem of calculating the current density through the sample of a material of finite width $L \gg a = 0.246~nm$ limited by two parallel electrodes is solved by calculating the transmission  probability $T(\varepsilon, p_{2}, V)$ in the presence of a given potential difference $V$ \cite{Vandecasteele:2010prb}:
\begin{equation}
\label{current_ldl}
j_{LDL} =  \frac{4e}{(2\pi\hbar)^2} \int dp_{2}  \int _{\varepsilon_{F}-eV}^{\varepsilon_{F}} d\varepsilon ~T(\varepsilon, p_{2}, V) .
\end{equation}
As before, an electric field is directed along the first axis of coordinates.
Therefore, the momentum component $p_{2}$ does not change its value during tunneling.
In spite of the fact that graphene is a gapless semiconductor, the presence of finite conserved values $p_2$ leads to the appearance of an energy gap with the width $\varDelta=2v_Fp_2$.
In the considered case of a vacuum initial state temperature and chemical potential are equal to zero, the Fermi energy $\varepsilon_{F}=0$ and there are no free carriers in the interelectrode space.
Under such conditions the process of transmission carriers can be carried out only by Zener-Klein tunneling.
The probability of such a tunneling in the WKB approximation is (taking into account the relationship of $\varepsilon$ and  $p_{2}$):
\begin{equation}
\label{probability}
T(\varepsilon, p_{2}, V) = T_{ZK} = \exp \Big(-\frac{\pi p_2^2 v_F L}{e \hbar V}\Big).
\end{equation}
In this case from (\ref{current_ldl}) and (\ref{probability}) one obtains \cite{Vandecasteele:2010prb}:
\begin{equation}
j_{LDL} =  2\frac{Ve^2}{\pi^3\hbar L} \sqrt{\frac{\pi^2eVL}{4\hbar v_F}} 
\times \big ( \text{erf}~ \Big{[}{\sqrt{\frac{\pi eVL}{4\hbar v_F}}}\Big{]}+\exp\Big{[}-\frac{\pi eVL}{4\hbar v_F}\Big{]}-1 \big ).
\label{current_t} 
\end{equation}

\section{Verification of two approaches \label{sect:comparison}}

First of all, let us note that the distribution (\ref{probability}) of carriers over the transverse momentum $p_2$ accurately reproduces the result (Fig.~\ref{fig:5}, right panel) obtained in the t-representation.

The basic problem is that the current density (\ref{currentcomp}) in the absence of dissipation and spatial boundaries depends on the time after switching on the field and increases continuously, which, obviously, is not observed in the experiment, where each value of the potential difference corresponds to its steady-state current density.
So, in the conditions of the real experiment \cite{Vandecasteele:2010prb} it is necessary to take into account availability of the electrodes, which limit the lifetime of the carriers.
From this point of view the process of carrier generation must permanently continue throughout the entire measurement process and be uniform in the area of the sample.
Knowing the strong anisotropy of the carrier spectrum, it is possible to assume in the first approximation that all of them move toward the electrode of the corresponding polarity with the velocity  $v_{F}$.
In this case, the average lifetime of carriers is $\tau=L/2v_F$ or $0.5 \times 10^{-12}~s$  for $L=1.0~\mu m$.
At the end of this time after switching on the field, the rate of generation of carriers becomes equal to the rate of their escape through the electrodes.
The steady-state current values will be constant and can be calculated in the both approaches (Table \ref{tab:table1}):
\begin{table}[h]
\caption{Comparison of calculated current density values for a sample with $L=1.0~\mu m$ in the range of potential differences from $0.1~V$ to $0.5~V$}\label{tab:table1}
\centering\small
\begin{tabular}{|c|c|c|} \hline
\parbox[c][3.5em]{8em}{\centering $V$ $[V]$ \\ ($E = V/L$ $[V/\mu m]$)}  & \parbox[c][3.5em]{10em}{\centering $j$ in x - representation $[\mu A/nm]$} &  \parbox[c][3.5em]{10em}{\centering $j$ in t - representation $[\mu A/nm]$}\cr \hline
 0.1 & 0.02904 & 0.02964 \cr \hline
 0.2 & 0.08344 & 0.08457 \cr \hline
 0.3 & 0.15435 & 0.15557 \cr \hline
 0.4 & 0.23861 & 0.24003 \cr \hline
 0.5 & 0.33440 & 0.33601 \cr \hline
 \end{tabular}
\end{table}

The results of these calculations coincide with high degree of accuracy in the considered range of parameters.
Indirectly it means good coincidence with experiment \cite{Vandecasteele:2010prb}.
The given list of values strictly corresponds to the law $I \sim V^{3/2} \sim E^{3/2}$.

\begin{figure}[h]
\centering
\sidecaption
\includegraphics[width=0.48\textwidth]{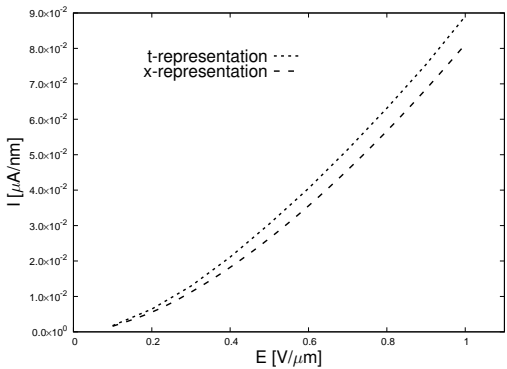}
\caption{Comparison of calculated current density values for a sample with $L=100~nm$ in the range of electrical field strengths $E$ from $0.1~V/\mu m$ to $1.0~V/\mu m$ (potential differences $V$ from $0.01~V$ to $0.1~V$).
\label{fig:4}}
\end{figure}

The kinetic method presented here can be a valid outside the framework of the applicability of the x - representation.
for example, in alternating electric fields in the region of sufficiently high frequencies.
An indirect confirmation of this is the difference in the results for a thinner (tenfold) sample with a correspondingly shorter carrier lifetime (Fig.~\ref{fig:4}).
In this case the difference reaches almost $10\%$.
Both approaches also predict superlinear growth of the current density.
But the exponent is already $1.70$ for both t - and x - representations.

\section{Conclusion}
\label{conclusion}

The kinetic theory presented here allows to describe carrier generation processes and the evolution of their distribution function in graphene under the action of an external time dependent electric field.
The possibilities of this method for a constant electric field with realistic parameters and the initial vacuum state was shown.
The I-V characteristics calculated in the kinetic approach are in good accordance with the predictions of the LDL approximation, which was verified in experiments with graphene samples.
A key advantage of the new approach is the ability to investigate the response not only to a constant electric field, but also to an electric field with an arbitrary time dependence and polarization.

Let us note that an essential problem would be direct a comparison of predictions of the nonperturbative kinetic theory in graphene \cite{bib_1, bib_2} with experiments in the optical range of the external field.

\subsection*{Acknowledgments}
The authors thank B.~Dora, D.~Kolesnikov, R.~Moessner, V.~Osipov and N.~Plakida  for useful discussion of the work.
This work was funded by RFBR according to the research project 18-07-00778 and the Polish NCN under grant number UMO-2014/15/B/ST2/03752.
D.B. acknowledges support by the MEPhI Academic Excellence programme under contract no. 02.a03.21.0005.

\end{document}